ENTROPIC ELASTICITY OF WEAKLY PERTURBED POLYMERS

Richard M. Neumann

The Energy Institute
The Pennsylvania State University
University Park, PA 16802
(January 22, 2001)

## Abstract

Taking into account the nonequivalence of fixed-force and fixed-length ensembles in the weak-force regime, equations of state are derived that describe the equilibrium extension or compression of an *ideal* Gaussian (freely jointed) polymer chain in response to an applied force in such a manner that the calculated unstretched *scalar* end-to-end separation is the random-coil size rather than zero, zero being the value predicted by the traditional Hookean entropy-spring formula. The entropy-spring model for a polymer chain is thereby modified so that for calculational purposes, the spring is of finite rather than zero unstretched length. These force laws, which result from accommodating the Brownian forces originating from the motion of the chain ends and opposing chain collapse, are shown to be consistent with observations from stretching experiments performed on single DNA molecules, wherein the measured extension approaches a non-zero limit as the external force is reduced. When used to describe single-chain dynamics, this approach yields a single-exponential relaxation expression for a short Gaussian chain (bead-spring dumbbell), which when initially compressed or extended relaxes into a state having the random-coil end separation, in agreement with the Rouse-model result. An equation is derived that describes the elongational response of a charged, tethered chain to a *weak* electric field (as might occur in electrophoresis), a calculation not feasible with the traditional approach. Finally, an expression for the entropic work required to bring the ends of a chain together, starting from the random-coil configuration, is derived and compared with the Hookean result.

## Introduction

Recently there has been considerable interest in the elastic properties of individual DNA and protein molecules.[1-6] The use of a variety of techniques including optical or magnetic tweezers, atomic-force cantilevers, microfibers, and hydrodynamic drag has permitted scientists to measure the retractive force as a function of the chain's end-to-end separation. Such experiments typically measure the elastic response of the chain in the strong-stretching regime where the chain becomes fully extended. There is general agreement with the prediction of the theory of ideal polymer chain elasticity in the moderate- and strong-stretching regimes in that the extension-versus-force behavior is linear until the end separation becomes an appreciable fraction of the fully extended chain length; at very high elongations, the behavior is best described by a Langevin-like functional dependence on the force.[1]

The elastic response of a single chain in the weak-stretching and compressive regimes is also of importance both from a theoretical as well as practical point of view. In this article, the term *weak-stretching* will refer to that force regime where chains have a scalar end-to-end separation lying between the random-coil value and four times this distance. *Random coil* will describe an idealized Gaussian chain at equilibrium in the absence of external forces. Forces operating in the weak-stretching regime are intermediate between those in the 1-to-50 piconewton (pN) range typically associated with DNA or protein stretching experiments and those of approximately $10^{-4}$ pN acting on single negative charges attached to a DNA segment in a 10 volt/cm electric field as might occur during electrophoresis. As an example, the force required to extend a 1-μm random coil of *idealized* DNA to 4 μm is approximately .03 pN. *Extension* will designate the average scalar end-to-end separation of the chain when this separation exceeds the random-coil value. In the weak-stretching regime, the distinction among various definitions of chain elongation (whereas immaterial in the moderate- and strong-stretching regimes) is critical. For example, the magnitude of the average of a vector end separation may be quite different from the average of the magnitude of the same vector quantity. *Compression* will refer to the condition of the chain when its scalar end-to-end separation is less than the random-coil value.

In the weak-stretching or compressive regime, the Brownian force, which arises from the fluctuation of the chain ends and acts in opposition to the Hookean force, can equal or surpass the Hookean force in magnitude. The inclusion of this Brownian force (as shown in a following section) in the equation of state for a polymer chain, in addition to eliminating certain paradoxical features of the traditional Hookean description, permits an ideal chain to be viewed as a spring of finite unstretched length, which can be compressed as well as extended from its random-coil configuration.

*Some Difficulties with Hooke's Law*

Eq. 1 describes the traditional Hooke's-law relationship between the vector force, $f$, and end-to-end separation, $r$:

$$\beta f = 2b^2 r \tag{1}$$

where $\beta = (k_B T)^{-1}$, $k_B$ is the Boltzmann constant, $T$ is the absolute temperature, and $b^2 = 3/2Na^2$. $N$ is the number of chain links, each of length $a$. Several authors have pointed out difficulties with the interpretation of Eq. 1 in the weak-stretching regime that are a consequence of the chain's remaining a microscopic system for finite $N$.[7-12] The most obvious problem is the contradiction between the existence of a random-coil chain whose ends are separated by a distance proportional to the square root of the number of links and a retractive-force law that suggests a zero end separation; this occurs when the force law is used without including the effects of Brownian motion. Moreover, Eq. 1 does not describe a chain's extension as a function of a constant, applied force, a situation which occurs when a chain, tethered at one end, is subjected to a weak external stretching force by means of fluid flow or an electric field; see Fig. 1. Eq. 1 also does not permit a physical description of the compression of an ideal Gaussian chain from its random-coil configuration to one having a smaller end separation.

Here, after a brief outline of the relevant theory, the following applications involving biopolymers will serve to illustrate the difference between the present approach and the Hookean description:

**(1)** The present force law explains the discrepancy sometimes observed between the classical theory and experiment when DNA-extension measurements are plotted versus external force, namely that the theoretical curve (used by the researchers) passes through the origin (zero extension, zero force) whereas the experimental points have an extension-intercept equal to the random-coil end separation of the molecule; see Fig. 1.[3,6,13]

**(2)** The extension of an ideal chain tethered at one end and carrying a charge subject to an electric field, a common occurrence when a DNA molecule becomes snagged on an obstacle during electrophoresis, is calculated for an idealized DNA segment as a function of the electric-field strength in the weak-stretching regime. This calculation is not possible in the Hookean description.

**(3)** A simple calculation is carried out showing that the entropic work required for a hypothetical enzyme to bring the ends of an idealized random coil of DNA together to form a loop or plasmid is a positive quantity, whereas the Hookean description results in a negative value for the work.

**(4)** A polymer chain, initially perturbed and then released from the constraint, is shown to relax into a final state having the random-coil end-to-end separation; this result can also be obtained from the Rouse model if the Brownian (white noise) force acting on the chain's ends is incorporated via the fluctuation-dissipation theorem.

## Nonequivalence of Ensembles

The ideal gas law, $\beta PV = n$, is a macroscopic equation of state describing equilibrium ideal-gas behavior when the number of molecules, $n$, is very large. Its existence is a consequence of the fact that the fluctuations in a state variable (volume for example) are vanishingly small when compared with the magnitude of that variable in the limit $n \to \infty$. For the ideal Gaussian polymer chain, the situation is different. By ideal Gaussian, we mean a chain that is located in a heat bath and consists of a large number of freely orienting volumeless segments, which do not interact with one another. The probability of occurrence of an end-to-end separation $r$ ($r = |\boldsymbol{r}|$) for such a chain is described by $r^2 P(r)$, where $P(r)$ is the normalized field-free end-to-end distribution function, $b^3 \pi^{-3/2} \exp(-b^2 r^2)$ for $r \ll Na$.[8] From $r^2 P(r)$ one may determine the standard deviation or dispersion ($\sigma_o$) in $r$ and compare it with the average value of $r$ for a freely orienting chain not subject to an external force. $\sigma_o^2 = <r^2>_o - (<r>_o)^2 = .23/b^2$. The absence of an external force is denoted by the subscript o. The result is:

$$\sigma_o / <r>_o = .42 \qquad (2)$$

It is apparent that the relative fluctuation is large and remains so regardless of the size of $N$. Similar behavior is observed for the relative fluctuation of force in a chain maintained at a fixed $r$.[11] As shown in the appendix, $\sigma/<r> \to 0$ in the limit $<r> \to Na$; the chain resembles a macroscopic system as the chain becomes fully extended.

Another characteristic feature of macroscopic systems is their equations of state are independent of the type of ensemble used in their derivation. Equivalent results for the ideal-gas law are obtained if the volume is held fixed and the pressure average calculated (volume ensemble) or if the pressure is held fixed and the volume average calculated (pressure ensemble).[14] For a polymer chain, $\boldsymbol{r}, \boldsymbol{f}, r$, and $f$ ($f = |\boldsymbol{f}|$) are state variables whose interrelationship is made explicit when deriving an equation of state. Three different ensembles are of particular interest in the weak-stretching regime. Here the results of calculations[8,9] for each ensemble are presented; the derivations are outlined in the appendix.

<u>In the constant $\boldsymbol{r}$ (displacement) ensemble</u>, the ends of the chain are fixed and an average force in the direction of $\boldsymbol{r}$ is determined[15-17]:

$$\beta<\boldsymbol{f}> = 2b^2 \boldsymbol{r} \qquad (3)$$
$$(\beta<f> = 2b^2 r)$$

<u>In the constant $\boldsymbol{f}$ (force) ensemble</u>, the chain ends are subject to a fixed force couple, and the average displacement (or other variable whose average value is desired) is calculated. This system may be realized by placing opposite charges on the ends of a chain and subjecting it to a constant electric field. The chain is assumed sufficiently long to ignore the interaction between the two charges. For the average displacement, the result is[8,18,19]:

$$\beta \boldsymbol{f} = 2b^2 <\boldsymbol{r}> \qquad (4)$$

Historically, Eqs. 3 and 4 have been regarded as describing a Hooke's-law dependence between force and displacement, their formal similarity taken as implicit evidence for an equivalence of ensembles, as in the macroscopic limit. Thus the <> symbol is often discarded, and the chain is regarded as analogous to a mechanical spring of zero unstretched length.

$|\langle r \rangle|$ is the magnitude of the average value of the projection of $r$ in the direction of $f$. It is illuminating to calculate $\langle |r| \rangle$ (i.e., $\langle r \rangle$) in the force ensemble and compare the result with Eq. 4[8]:

$$\langle r \rangle = \langle r \rangle_o [1 + v^2/12 + O(v^4)] \quad (5)$$

$v = \beta f/b$; the equation is accurate for $v < 3$. An expression for $\langle r \rangle$ valid for larger $v$ is given by Eq. 13 in the appendix. Under the influence of a small force, the chain maintains a nearly constant average end-to-end separation, $\langle r \rangle_o$, (the random-coil size) while "rotating" analogous to an electric dipole; hence $\langle r \rangle$ increases linearly, beginning at zero, with the force. That $|\langle r \rangle|$ and $\langle r \rangle$ differ in their physical interpretation has been a long-standing source of confusion. If the elongation measurements in a fixed-force stretching experiment are averaged to obtain $|\langle r \rangle|$, then ideally, the system would obey Hooke's law in the weak-stretching regime. As indicated, this is not true if the experiment is analyzed in terms of $\langle r \rangle$ or $\langle |x| \rangle$.

In the constant $r$ (length) ensemble, the scalar separation between the chain ends is maintained fixed while the orientation of $r$ is permitted to vary. An "imaginary rod" of length $r$ is inserted between the ends of the chain to fix the length. The average tension, $f$, in the rod may then be determined[9]:

$$\beta \langle f \rangle = 2(b^2 r - 1/r) \quad (6)$$

Eq. 6 is more realistic than Eq. 1 in the sense that it permits the existence of a randomly coiled chain without the need for ad hoc postulates to prevent chain collapse. The force vanishes when $r = r_p = 1/b$, the most probable value of $r$, rather than when $r = 0$. Here the ideal chain behaves as a spring of *finite* unstretched length, which may be compressed as well as extended; a negative value for $\langle f \rangle$ corresponds to compression. This equation of state is also implicit in Flory's derivation of the molecular expansion factor, $\alpha$, although in a different context.[20] The $2/\beta r$ repulsive force, whose origin is in the Brownian fluctuation of the chain's ends, was first introduced by the author as an intuitive explanation of why an ideal chain does not collapse.[21]

Fig. 2 shows the extension versus force behavior predicted by three different equations of state. As expected, in the moderate-stretching region, the three merge into Eq. 1. When the dimensionless force exceeds 7.5 (corresponding to $r > 4r_p$ in the length ensemble), the variation among the different calculated extensions is less than five percent. This is the reason for the arbitrary definition of the weak-stretching regime given in the introduction. In this regime, the appropriate equation of state depends on the particular physical environment of the chain. In fact, Eq. 3 is not applicable at all because its derivation is predicated on "fixed" ends. As shown in the appendix, any finite orientational fluctuation in $r$ causes the introduction of the $1/r$ term.

## Elastic Behavior of Biopolymers Subjected to Perturbations

*DNA-Stretching Experiments and Hooke's Law*

Fig. 1 shows the extension of a DNA molecule, visualized with fluorescence microscopy, as a function of the flow velocity of the solvent surrounding it.[6] The molecule was held stationary against the flow by means of a microsphere attached to one end; the microsphere, in turn, was secured by means of optical trapping. The measure of extension used by the authors was the average maximum visual elongation in the direction of flow. Here it is assumed that this quantity is proportional to $\langle |x| \rangle$, $x$ being the component of $r$ the flow direction. $\langle |x| \rangle$ may be approximated by $\langle r \rangle$; see figure caption.

Because the experiment was conducted in a fixed-force mode, the constant-force ensemble result, Eq. 13, is applicable if one regards the fluid flow as a source of potential[12]. The extension-versus-force (the force is assumed to be proportional to flow velocity) curve is consistent with Eq. 13 (see Figs. 1 and 2) in that the extension expected (by the authors) in the limit of zero force is the radius of gyration, $R_g$, and the behavior becomes Hookean in the moderate stretching regime. Since $R_g$ was not provided by the authors and had to be estimated, and given the general paucity of data in the weak-stretching region, no attempt was made to fit the data to Eq. 13. It should be emphasized that the definition of extension used by the authors to analyze the experiment requires that the zero-force extension limit be non-zero, thereby precluding a Hooke's-law description. In their analysis[6], the authors used an equation derived for a worm-like chain (WLC) because of its applicability in describing DNA in the moderate-to-strong stretching region. Because this equation reduces to one for a simple Hookean dumbbell (Eq. 1) in the weak-stretching region, their theoretical description is inaccurate in this region, as shown by the solid line in Fig. 1. (The WLC in this context is discussed in Ref. 12.) It is also assumed by the authors (and here) that the force acts solely on the bead at the end of the chain; in reality, the viscous drag force is cumulative and acts most strongly on the tethered end and least strongly on the free end.

*Electrophoresis*

Electrophoresis is a common technique wherein polymer molecules carrying a net charge may be separated according to their molecular weights. Individual chains migrate through a matrix under the influence of a weak electric field. The matrix may consist of a gel or it may be a micro-fabricated array.[1] For DNA, the gel-electrophoresis technique works best using electric fields of less than 10 volts/cm with relatively short molecules of fewer than 20 kilobases.[1,22] For a longer chain or stronger field, the molecule elongates if an end becomes temporarily entangled in the matrix; the chain subsequently relaxes back to its random-coil size as the end slides free. Cycles of expansion and contraction can cause the technique to become unreliable. Thus, to reduce the length of time required for an electrophoretic separation, it is sometimes advantageous to use the maximum electric-field strength possible that does not initiate elongation of the chain.

From this consideration as well as a desire to demonstrate the applicability of our force equation, we shall calculate the relative increase in $<r>$ from its random coil length as a function of net charge, electric-field strength, and $<r>_o$ for an *idealized* strand of DNA in order to predict the onset of chain extension. We neglect the effect of counterions present in the surrounding solution and other complications which would occur in an actual experiment. The constant-force-ensemble result, Eq. 5, applies if (1) the matrix entanglement described above is of sufficient duration to permit an equilibrium description and (2) the medium is sufficiently sparse (large pores) to ensure unrestricted motion of the individual chain segments. The more accurate expression from the appendix, Eq. 13, should be used if the relative extension is greater than 100 percent. From Eq. 5, the following relationship between percent increase in extension, $h$, and electric-field strength may be derived for $T = 300$ K:

$$h = 9.8 \times 10^3 (ZE<r>_o)^2 \qquad (7)$$

where $h = 100[<r> - <r>_o]/<r>_o$, $E$ is the electric-field strength in volts/m, and $Z$ is the integral multiple of the electron charge; $|Z| = 1$ for a single electron or proton. It has also been assumed that the net charge present on the strand of DNA is located on the free end. In reality, the charge is distributed along the length of the entire chain. Thus, only the chain segment between the tethered end and the location of the next charge (proceeding toward the free end) experiences the full force, $EZq$, where $q$ is the electron charge. For this reason, the average charge should be used in the above expression, which would then describe a strand carrying a charge of $2Zq$. As a numerical example, consider a yeast chromosome of approximately one-million base pairs having an equilibrium coil size, $<r>_o$, equal to 5 μm.[1] From Eq. 7, for such a molecule to be stretched by ten percent, $ZE = 6300$. If the chromosome carries a net charge of $500q$ ($Z = 250$), the required electric-field strength would be approximately .25 volts/cm.

*Random-Coil Compression*

A random coil may be compressed as well as stretched. A biological illustration of the use of Eq. 6 in chain compression would be the calculation of the entropic work, $W$, required for an enzyme to bring the ends of a chain from the random-coil separation, $1/b$, to a distance $a$ from one another, as would occur in the formation of a loop of DNA (plasmid). In this idealization, the enzyme "grasps" the chain ends in such a fashion as to approximate the constant-length criterion required for the application of Eq. 6. $W = \int_{1/b}^{a} f \, dr = (\ln N + 3/2N - 1.4)/\beta$; it is assumed that the compression occurs slowly to ensure equilibrium and that excluded volume is absent. As the force is directed along $r$; $f \cdot dr = f \, dr$. In the limit of large $N$, $W = (\ln N)/\beta$. Finally, for comparison, the work obtained from simply applying Eq. 1 (instead of Eq. 6) in the limit of large N is: $W = -1/\beta$.

*Random-Coil Perturbation with Subsequent Relaxation*

The traditional description of the relaxation behavior of a single polymer chain is based on the Rouse model,[23,24] which for a two-bead spring dumbbell is a Langevin equation: $-\zeta dr/dt = \partial U/\partial r - f(t)$. $\zeta$ is the friction factor for a bead; $U$ the spring potential energy; and $f(t)$ the random (noise) Brownian-force vector acting on a given bead. Frequently when solving this type of equation, particularly for chains at equilibrium, $f(t)$ is averaged to zero through coarse-graining, resulting in the collapse of the spring to $r = r = 0$ for $t \to \infty$. Such a model is unrealistic when considering the behavior of a random-coil segment of DNA that is perturbed (either by extension or compression) from its equilibrium configuration and permitted to return to this state. The perturbation could occur naturally in a living organism, such as when stresses are transmitted from a cell's skeletal structure inward to the nucleus and chromosomes, or be a result of laboratory manipulation as in DNA-relaxation experiments[25].

To illustrate the use of Eq. 6 in describing polymer-relaxation behavior, a chain of just sufficient length to obey Gaussian statistics is chosen rather than a superchain consisting of many Gaussian subchains, as in the complete Rouse model. This simplification results in a single relaxation mode[23], rather than a spectrum of relaxation modes and times. The chain is tethered at one end, contains a bead at the other end, and is immersed in a viscous medium. The chain's frictional drag is regarded as occurring solely at the site of the bead; the rest of the chain serves merely as a spring, as in a dumbbell. The equation of motion is given by

$$-(\zeta \beta) dr/dt = 2(b^2 r - 1/r). \tag{8}$$

Integration of this equation, with the initial condition that $r_o = c/b$ ($c \geq 0$), yields:

$$(br)^2 = 1 + (c^2 - 1)\exp(-t/\tau), \tag{9}$$

where the relaxation time, $\tau$, is equal to $\zeta\beta/4b^2$. For $c > 1$, the initial state is one of extension; for $c < 1$, it is one of compression. As anticipated, $r \to 1/b$ (not 0) as $t \to \infty$. It is readily shown that for long chains whose ends are initially "touching" ($r = 0$ at $t = 0$), $r^2 = 4Dt$ for $t \ll \tau$, which is essentially the well-known Einstein formula, where $D = 1/\beta\zeta$. Thus for such a chain, the end points diffuse away from one another, unaffected by the Hookean spring connecting them, provided $t \ll \tau$.

Eq. 9 is, in fact, consistent with experimental observations[3] where the relaxational behavior of individual DNA molecules was observed by fluorescence microscopy. Highly stretched molecules were released at one end, and the extension was measured as a function of time. For $r < .3Na$, the $r^2$-versus-time behavior could be described by means of exponential decay with a single relaxation time.[3]

Finally, we wish to emphasize that an expression equivalent to Eq. 9 can be obtained by solving the Langevin equation, making use of the fluctuation-dissipation formula, $\langle f(t)f(t')\rangle =$ $(2\zeta/\beta)\delta(t - t')$[24]. The expression for $\langle f(t)f(t')\rangle$ and the repulsive force $2/\beta r$ are equivalent representations of the thermal forces acting on a given bead.

## Conclusion

The behavior of a Gaussian chain in the weak-stretching region is a direct result of the chain's remaining a microscopic thermodynamic system. Eq. 3, although the correct equation for interpreting current DNA- and protein-stretching experiments in the moderate-stretching regime, is in reality an asymptotic limit of the constant-force or constant-length ensemble equation of state. The force equation obtained from the constant-length ensemble, Eq. 6, encompasses compression as well as extension; in extension it is similar to Eq. 13, as shown in Fig. 2. The origin of this similarity lies in the fact that the constant-length and constant-force ensembles both involve calculating averages over all allowed orientations of *r* and hence implicitly include Brownian forces, which arise from the motion of the chain's ends and which oppose chain collapse.

The use of a constant-force ensemble result (Eq. 5 or 13) permits the calculation of the increase in $\langle r\rangle$ as a function of the external force in the initial phase of deformation for a polymer such as DNA. This calculation is not possible using the traditional Hooke's-law expression, which accounts for the discrepancy between theory and experiment observed in plots of chain extension versus force where the initial data points do not lie on a straight line[6,13], as shown in Fig. 1. It should be emphasized that Hooke's law would apply in the weak-stretching regime if a fixed-force experiment were analyzed using Eqs. 4 or 11; however, the preferred measure of extension is usually $\langle r\rangle$ rather than $|\langle r\rangle|$[3,6,13,25], thereby vitiating a Hookean treatment.

Presently, a considerable effort is underway to increase the speed and resolution of the electrophoresis technique as it is an essential part of many DNA-sequencing methodologies. Understanding the elastic properties of polymer chains is crucial to the design of improved electrophoretic chambers. Researchers frequently use an analysis based on Eqs. 3 and 4 in calculating the response of DNA fragments (chains) to a uniform electric field for all degrees of stretching.[1,26-29] As noted earlier, if the desired measure of chain elongation is $\langle r\rangle$ rather than $|\langle r\rangle|$, a significant error may result.

Small changes in *r* for chromosomes or other random-coil-like structures inside a cell that result from mechanical force transmission from elsewhere in the cell or from the action of local enzymes or motor proteins cannot be described using Eqs. 1, 3, or 4, particularly if the stress causes a decrease in *r*. The tube model[24], introduced to explain single-chain behavior in a rubber-like matrix environment, invokes a "hypothetical tensile force" similar to $2k_BT/r$ to prevent the collapse of a "primitive" chain. Since the tube model is sometimes used to describe single DNA chains in a gel matrix during electrophoresis, a non-Hookean approach is also implied here.

Finally, whereas our results are rigorous for ideal chains, they should be viewed as conceptual stepping stones when considering the effects of excluded volume and other interactions. To the best of our knowledge, the nonequivalence of ensembles remains unrecognized in *ideal* polymer chains. Some workers have examined this problem in real chains, however.[7,10,11,30]

## Appendix

<u>$\langle r\rangle$ Fluctuations at High Extension</u>
The following relationship[19] is accurate for calculating $\sigma$ at fixed force, *f*, for $v > 8$: $\sigma^2 = \beta^{-1}(\partial\langle r\rangle/\partial f)_T$. In the moderate-to-strong stretching regime, $\langle r\rangle$ may be replaced by *r* from Eq. 1 yielding $\sigma^2 = 1/2b^2$. Comparing $\sigma$ with *r*: $(\sigma/r)^2 = 1/\beta rf$. As the chain is extended, *rf* increases causing a decrease in the relative fluctuation in *r*. As noted earlier, at high extensions Eq. 1 must be replaced by an expression involving the Langevin function, resulting in $f \to \infty$ and $\sigma/r \to 0$ in the limit $r \to Na$. Similarly, it may be shown that for a fixed finite force, $\sigma/r \to 0$ in the limit $N \to \infty$. This behavior is consistent with the merging of the three equations of state as depicted in Fig. 1, as a consequence of $v \to \infty$ as $N \to \infty$.

Constant-Displacement Ensemble

The average vector force resulting from a fixed displacement, $r$, is obtained from the chain entropy by means of the Boltzmann relationship between $S$ and $P(r)$:

$$<f> = -T(\partial S/\partial r)_T = 2k_B Tb^2 r, \qquad (10)$$

where $S = k_B \ln[P(r)]$.

Constant-Force Ensemble

The average properties of a variable $Q$ are calculated from $<Q> = H/J$. Define $A(r) = P(r)\exp(\beta f \cdot r)$. Then $H = \int Q\, A(r)\mathrm{d}w$ and $J = \int A(r)\mathrm{d}w$, where $\mathrm{d}w = r^2 \sin\theta\, \mathrm{d}\theta\, \mathrm{d}\phi\, \mathrm{d}r$. $\theta$ is the angle between $f$ and $r$; $x$ is the component of $r$ in the direction of $f$, $x = (r \cdot f/f) = r\cos\theta$. The following averages may be derived:

$$<r> = <x>f/f \quad \text{and} \quad |<r>| = <x> = v/2b, \qquad (11)$$

where $v = \beta f/b$. If the $\exp(\beta f \cdot r)$ term inside each of the integrals, $H$ and $J$, is expanded in a power series, one obtains

$$<r> = <r>_o [1 + v^2/12 + O(v^4)]. \qquad (12)$$

This slowly converging series is cumbersome for $v > 3$ and may be replaced with the "exact" result, which is valid in the region were the chain statistics remain Gaussian, i.e. for $r \ll Na$:

$$<r> = (<r>_o/2)[\exp(-v^2/4) + (v + 2/v) \int_0^{v/2} \exp(-t^2)\, \mathrm{d}t] \qquad (13)$$

It can be shown that Eq. 13 simplifies to Eq. 1 in the limit of moderate forces. Similarly, in the limit $f \to 0$, $<r> = <r>_o$.

Constant-Length Ensemble

Here a constant (scalar) length is established by attaching each end of a Gaussian chain to an end of a rigid rod of length $r$. The rod, in turn, is subject to a force that biases its orientation. A means of accomplishing this is to attach one end of the rod to a fixed point (about which it may swivel freely) and placing a charge ($Zq$) on the other end subject to an electric field. The equation of state is determined by calculating the average tension in the rod as a function of $r$. The partition function $\Delta$ for the chain is given by

$$\Delta = \int \mathrm{d}L\, P(L)\exp(\beta F \cdot L)\, \delta(L - r), \qquad (14)$$

where $F = ZqE$, $\mathrm{d}L = L^2 \sin(\theta)\, \mathrm{d}\theta\, \mathrm{d}\phi\, \mathrm{d}L$ and $\delta$ is the Dirac delta function. The average tension, $<f>$, between the chain ends follows from

$$<f> = -(\partial \ln\Delta/\partial r)/\beta = (2/\beta)(b^2 r - 1/r) - (ZqE)L_g(Zq\beta Er), \qquad (15)$$

where $L_g$ denotes the Langevin function; $L_g(x) = \coth(x) - 1/x$. A minus sign was placed in front of the partial derivative so that the Hookean term in the force expression would be positive, in keeping with convention. The first term in the above equation yields Eq. 6; the significance of the second term is that by increasing the electric field strength, *E*, one may suppress the orientational fluctuation of the rod (and hence *r*) to any degree desired. Thus in the limit $E \to \infty$, the constant-length ensemble degenerates into a constant-displacement (*r*) ensemble without the loss of the $1/r$ term in the equation of state. An electrical interaction was arbitrarily introduced to illustrate this point; another force could have sufficed. In the examples where Eq. 6 is used, no orienting interaction is involved, and the second term in Eq. 15 is ignored.

## Acknowledgments

I thank Harold H. Schobert and George Fleck for many stimulating conversations and much encouragement. I am grateful to J. H. Weiner and Pierre LeBreton for critically reading an early draft of this manuscript.

## References

1. R. H. Austin, J. P. Brody, E. C. Cox, T. Duke, and W. Volkmuth, *Physics Today* **50**, 32 (1997).
2. H. Yin, M. D. Wang, K. Svoboda, R. Landick, S. M. Block, and J. Gelles, *Science* **270**, 1653 (1995).
3. T. T. Perkins, D. E. Smith, and S. Chu, *ibid.* **276**, 2016 (1997). [Note #26 contains information on relaxation behavior.]
4. M. D. Wang, H. Yin, R. Landick, J. Gelles, and S. M. Block, *Biophys. J.* **72**, 1335 (1997).
5. C. Bustamante, J. F. Marko, E. D. Siggia, and S. Smith, *Science* **265**, 1599 (1994).
6. T. T. Perkins, D. E. Smith, R. G. Larson, and S. Chu, *ibid.* **268**, 83 (1995).
7. R. A. Guyer and J. A. Y. Johnson, *Phys. Rev. A* **32**, 3661 (1985).
8. R. M. Neumann, *ibid*. **31**, 3516 (1985).
9. R. M. Neumann, *ibid*. **34**, 3486 (1986).
10. D. H. Berman and J. H. Weiner, *J. Chem. Phys.* **83**, 1311 (1985).
11. J. H. Weiner and D. H. Berman, *J. Polymer Sci., Polymer Phys. Ed.,* **24**, 389 (1986).
12. R. M. Neumann, *J. Chem. Phys.* **110**, 7513 (1999).
13. S. B. Smith and A. J. Bendich, *Biopolymers* **29**, 1167 (1990).
14. A. Münster, *Statistical Thermodynamics*, (Springer-Verlag, New York, First Engl. Ed.,1969), Chapter IV.
15. P. J. Flory, "Polymer chain elasticity" *Proc. R. Soc. London, Ser. A* **351**, 351-360 (1976).
16. H. M. James and E. Guth, *J. Chem. Phys.* **11**, 455 (1943).
17. L. R. Treloar, *The Physics of Rubber Elasticity* (Oxford, London, 1958).
18. I. Webman, J. L. Lebowitz, and M. H. Kalos, *Phys. Rev. A* **23**, 316 (1981).
19. T. L Hill, *An Introduction to Statistical Thermodynamics* (Addison-Wesley, Reading, MA, 1962), pp. 217-218.
20. P. J. Flory, *Principles of Polymer Chemistry* (Cornell, Ithaca, 1971), p. 600.
21. R. M. Neumann, *J. Chem. Phys.* **66**, 870 (1977).
22. J. Noolandi in "Theory of DNA Gel Electrophoresis": *Advances in Electrophoresis* A. Chrambach et al., Eds. (VCH, Weinheim, New York, Cambridge, 1992) Vol. **5**, pp. 1-56.
23. P. G. deGennes, *Scaling Concepts in Polymer Physics* (Cornell, Ithaca, 1979), pp. 167-180.
24. M. Doi and S. F. Edwards, *The Theory of Polymer Dynamics* (Oxford, New York, 1988).
25. T. T. Perkins, S. R. Quake, D. E. Smith, and S. Chu, *Science* **264**, 822 (1994).
26. T. Duke, G. Monnelly, R. H. Austin, and E. C. Cox, *Electrophoresis* **18**, 17 (1997).
27. W. D. Volkmuth, T. Duke, M. C. Wu, and R. H. Austin, *Phys. Rev. Lett.* **72**, 2117 (1994).
28. O. B. Bakajin, T. A. J. Duke, C. F. Chou, S. S. Chan, R. H. Austin, and E. C. Cox, *Ibid*. **80**, 2737 (1998).
29. W. D. Volkmuth and R. H. Austin, *Nature* **358**, 600 (1992).
30. J. T. Titantah, C. Pierleoni, and J. -P. Ryckaert, *Phys. Rev. E* **60**, 7010 (1999).

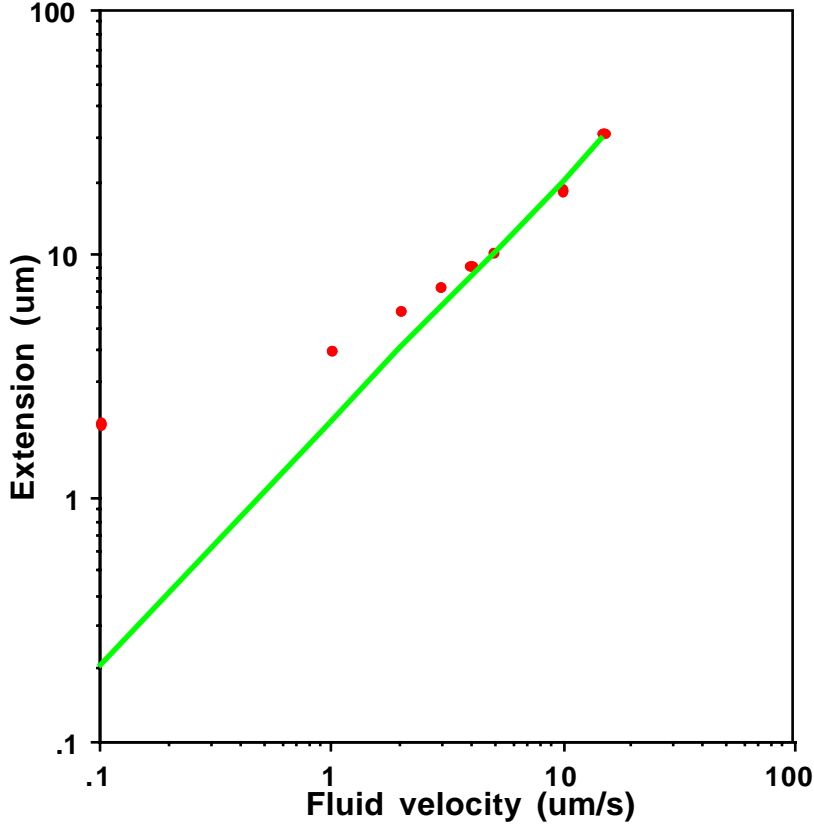

**Fig. 1.** (•) depict extension measurements as a function of the fluid velocity for an individual, fluorescently labeled DNA molecule using data in the weak/moderate-stretching regime adapted from Fig. 1 of Ref. 6. The initial extension here is based on an estimate for $R_g$, as the authors provided no value for this quantity. The solid line (—), also borrowed from Ref. 6, illustrates Hookean behavior "predicted" from a dumbbell model based on Eq. 1. In the experiment, a series of "maximum visual extensions in the direction of flow" for a single fluctuating molecule (at a fixed solvent flow rate in the $x$ direction) was measured. The individual measurements, taken at approximately 1-second intervals, were averaged to obtain the reported extension, which corresponds to $<|x|>$ rather than $<r>$. A constant-force ensemble calculation may be used to relate these averages:

$$<|x|> = <r> - (<r>_o/v)\int_0^{v/2} \exp(-t^2)\,dt\,. \tag{16}$$

Because $<|x|> \rightarrow <r>$ with increasing force, a distinction between these two measures of extension is not necessary for the present purpose. As expected, in the limit of zero force (fluid velocity), $<|x|>_o = <r>_o/2 \sim R_g$.

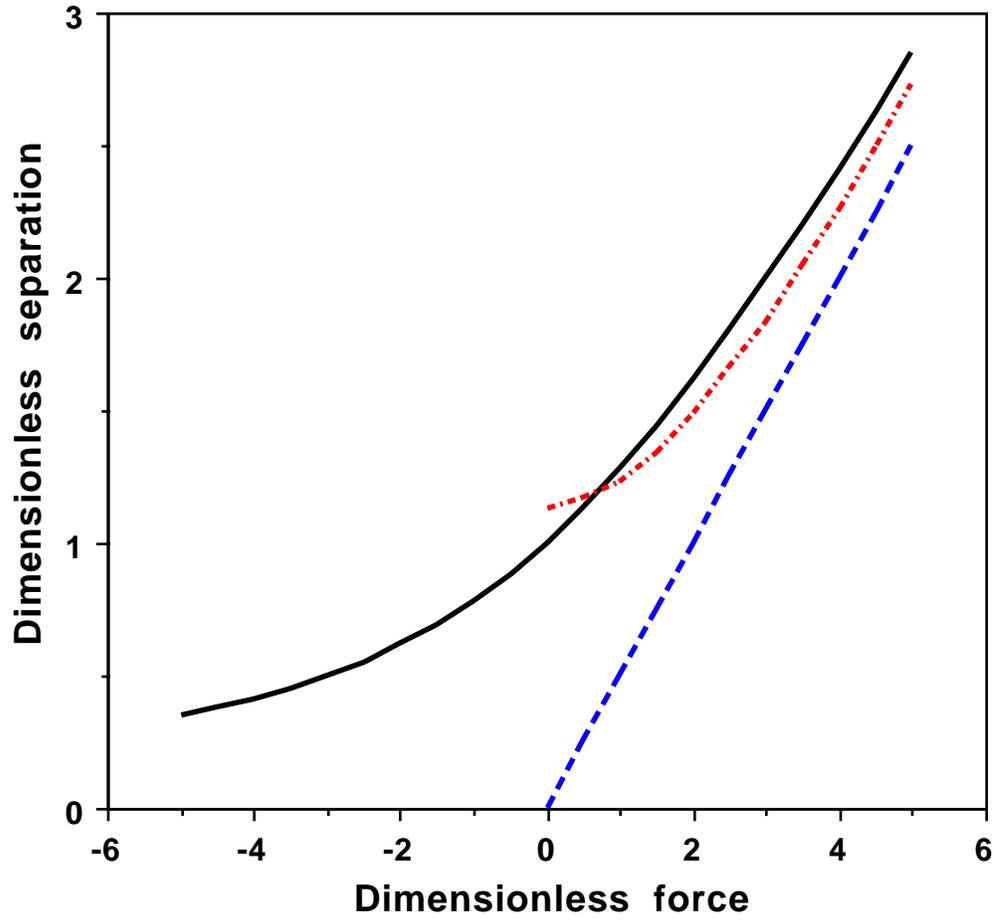

**Fig. 2.** The theoretical dimensionless end-to-end separation is shown as a function of the dimensionless force for the three ensembles. (——) depicts $br$ versus $\beta \langle f \rangle/b$ from Eq. 6 of the length ensemble. (—— •) depicts $b\langle r \rangle$ versus $\beta f/b$ from Eq. 13 of the force ensemble. (- - -) depicts both $br$ versus $\beta \langle f \rangle/b$ from Eq. 3 of the displacement ensemble and $b\langle x \rangle$ versus $\beta f/b$ from Eq. 11 of the force ensemble.